# Distributed Stochastic ACOPF Based on Consensus ADMM and Scenario Reduction

Shan Yang
School of System Science and Engineering
Sun-Yat Sen University
Guangzhou, China
yangsh237@mail2.sysu.edu.cn

Yongli Zhu*
School of System Science and Engineering
Sun-Yat Sen University
Guangzhou, China
yzhu16@alum.utk.edu (corresponding author)

*Abstract*—This paper presents a Consensus ADMM-based modeling and solving approach for the stochastic ACOPF. The proposed optimization model considers the load forecasting uncertainty and its induced load-shedding cost via Monte Carlo sampling. The sampled scenarios are reduced using a clustering method combined with simultaneous backward reduction techniques to reduce computational complexity. The proposed approach is tested on two IEEE systems, achieving more than 2% cost reduction and more than 15 times lower reliability index in stochastic load settings compared to the baseline approach.

## I. INTRODUCTION

AC Optimal Power Flow (ACOPF) is a fundamental problem in power systems that is widely used in system planning and operation, aiming to optimize a pre-defined objective function of the power grid under specified constraints. Interconnected power systems are a group of regional power grids connected via a few long-distance tie-lines. Due to the administrative barrier, economic interest, and data privacy concerns, the convention-centralized ACOPF scheme might not be welcomed by all regional entities of the interconnected power systems. Therefore, distributed optimal power flow (OPF) has been studied [1]-[4] and plays a key role as the interconnected power systems become larger and larger.

ADMM (Alternating Direction Method of Multipliers) is a famous algorithmic framework for decomposable optimization problems. The vanilla ADMM works in an alternative rather than parallel manner and requires regional information change when applied to the interconnected systems' OPF [5]; hence, it is not a suitable solution for regional data privacy. Consensus ADMM [6], as a parallelizable variant of ADMM, has been widely used in decomposable problems where variables can be divided into groups and associated via a group of common variables (called *consensus variables*). It has been successfully applied in machine learning and signal processing, e.g., solving large-scale SVM (support vector machine) problems [6] and multi-agent learning [7]. It has also been applied in the standard ACOPF to relieve the original problem's nonconvexity [8].

Meanwhile, incorporating load-induced stochasticity into the OPF model is also crucial due to the real-world load fluctuations and inherent uncertainties in load forecasting [9]. A stochastic modeling approach helps the power system operator "foresee and prepare" for a range of possible scenarios and provides slightly conservative but more resilient solutions to balance the trade-offs between reliability and cost-effectiveness. Different stochastic optimization methodologies can lead to different reformulations [10]-[12], which usually enlarge the problem dimension (in terms of decision variables and/or constraints). This dimension explosion issue may increase the solution time and exacerbate the original problem's convergence. Thus, scenario reduction techniques are needed, which shrink the original set of scenarios into a smaller "delegate set". For example, spectral clustering has been utilized for substation load data [13]. In [14], a clustering method is adopted to extract typical operation scenarios. In [15], a scenario reduction solution is developed for power market trading.

In this paper, we establish a distributed stochastic ACOPF model that incorporates the load (forecasting) uncertainty and penalizes such uncertainty by adding an extra cost term in the objective function. The model is then solved by using consensus ADMM after applying a scenario reduction procedure. The proposed approach can find an OPF solution with improved system reliability under stochastic load scenarios.

Regarding the remaining sections, Section II describes the stochastic ACOPF model proposed in this paper. Section III introduces the basic idea of the consensus ADMM and the K-means-based scenario reduction approach. Section IV presents case studies on two IEEE systems based on the proposed model and compares the results with the base case regarding the objective value and the slack-power shortage rate. Conclusions and future steps are given in the final section.

## II. A STOCHASTIC REFORMULATION OF ACOPF

### A. A Standard Version of ACOPF

A standard ACOPF model can be described by Eq. (1):

$$\min_{V,\theta,P,Q} f(P) = \sum_{i=1}^{n_g} \left( a_i P_i^2 + b_i P_i + c_i \right) \quad (1a)$$

$$\text{s.t.} \quad P_i^{\min} \leq P_i \leq P_i^{\max}, \quad i = 1,...,n_g \quad (1b)$$

$$Q_i^{\min} \leq Q_i \leq Q_i^{\max}, \quad i = 1,...,n_g \quad (1c)$$

$$V_i^{\min} \leq |V_i| \leq V_i^{\max}, \quad i = 1,...,n_b \quad (1d)$$

$$\theta_i^{\min} \leq \theta_i \leq \theta_i^{\max}, \quad i = 1,...,n_b \quad (1e)$$

This work is supported by "Scientific Computing and System Modeling: Teaching Reform Project" (45000-51200002)

$$P_i - P_i^d - \sum_{j=1}^{n_b} V_i V_j (G_{ij} \cos(\theta_i - \theta_j) + B_{ij} \sin(\theta_i - \theta_j)) = 0, \quad i = 1,...,n_b \quad (1f)$$

$$Q_i - Q_i^d - \sum_{j=1}^{n_b} V_i V_j (G_{ij} \sin(\theta_i - \theta_j) - B_{ij} \cos(\theta_i - \theta_j)) = 0, \quad i = 1,...,n_b \quad (1g)$$

Here, $n_b$ is the total number of buses, $n_g$ is the total number of generators. $(a_i, b_i, c_i)$ are the cost parameters of the $i$-th generator. $(V_i, \theta_i, P_i, Q_i)$ are the voltage magnitude, the voltage angle, the active power output, and the reactive power output of the generator at bus $i$. $(P_i^d, Q_i^d)$ are the active and reactive load at bus $i$. $G_{ij}$ is the conductance between bus $i$ and bus $j$. $B_{ij}$ is the susceptance between bus $i$ and bus $j$.

### B. A Proposed Stochastic Version of ACOPF

The stochastic ACOPF model proposed in this paper is based on the following stochastic programming model in Eq. (2), where $\omega$ stands for a random variable, $x$ is the decision variable, and "**E**" is the expectation cooperator. $f_0$ and $\mathcal{X}$ are, respectively, the objective function and the feasible region. This type of problem tries to minimize the expectation of $f_0$ within the feasible region.

$$\min \quad F_0(x) = \mathbf{E} f_0(x, \omega) \quad (2a)$$
$$\text{s.t.} \quad x \in \mathcal{X} \quad (2b)$$

In this paper, the demand at each load bus is deemed as the random variable. $M$ scenarios of load demand are considered, each associated with a probability $\mu_s$ ($s=1,...,M$). The goal is to find an optimal solution that can accommodate all scenarios in a probabilistic sense. More specifically, compared to the standard ACOPF, we make the following modifications:

- Modified Objective Function.

Our objective function is to minimize the total operating cost, considering both generation cost and loss of load cost:

$$\min_{V,\theta,P,Q,\Delta P_{i,s}^d} c \sum_{s=1}^{M} \mu_s \sum_{i=1}^{n_b} \Delta P_{i,s}^d + \sum_{i=1}^{n_g} \left( a_2 P_i^2 + a_1 P_i + a_0 \right) \quad (3)$$

where $c$ is the unit penalty cost for load shedding, $\mu_s$ is the probability of scenario $s$, and $\Delta P_{i,s}^d$ is the loss of load at bus $i$ in scenario $s$, which is a newly added decision variable.

- Loss of Load Constraints

The loss of load at each bus $i$ in each scenario $s$ should not exceed its total load and should be nonnegative:

$$0 \leq \Delta P_{i,s}^d \leq P_{i,s}^d, \quad s = 1,...,M \quad (4)$$

- Modified Active Power Constraints

For each bus $i$ and each scenario $s$, the active power surplus should be nonnegative when considering load shedding:

$$\Delta P_{i,s}^d + P_i - P_{i,s}^d - \sum_{j=1}^{n_b} V_i V_j \left( G_{ij} \cos(\theta_i - \theta_j) + B_{ij} \sin(\theta_i - \theta_j) \right) \geq 0 \quad (5)$$

These modifications ensure that the proposed stochastic ACOPF model accommodates the impact of the load forecast errors while minimizing the total operational costs.

Finally, the proposed stochastic ACOPF model is summarized in Eq. (6).

$$\min_{V,\theta,P,Q,\Delta P_{i,s}^d} c \sum_{s=1}^{M} \mu_s \sum_{i=1}^{n_b} \Delta P_{i,s}^d + \sum_{i=1}^{n_g} \left( a_2 P_i^2 + a_1 P_i + a_0 \right) \quad (6a)$$

$$\text{s.t.} \quad P_i^{\min} \leq P_i \leq P_i^{\max}, \quad i = 1,...,n_g \quad (6b)$$

$$Q_i^{\min} \leq Q_i \leq Q_i^{\max}, \quad i = 1,...,n_g \quad (6c)$$

$$V_i^{\min} \leq |V_i| \leq V_i^{\max}, \quad i = 1,...,n_b \quad (6d)$$

$$\theta_i^{\min} \leq \theta_i \leq \theta_i^{\max}, \quad i = 1,...,n_b \quad (6e)$$

$$0 \leq \Delta P_{i,s}^d \leq P_{i,s}^d, \quad i = 1,...,n_b, \quad s = 1,...,M \quad (6f)$$

$$Q_i - Q_i^d - \sum_{j=1}^{n_b} V_i V_j (G_{ij} \sin(\theta_i - \theta_j) - B_{ij} \cos(\theta_i - \theta_j)) = 0 \quad (6g)$$

$$\Delta P_{i,s}^d + P_i - P_{i,s}^d - \sum_{j=1}^{n_b} V_i V_j \left( G_{ij} \cos(\theta_i - \theta_j) + B_{ij} \sin(\theta_i - \theta_j) \right) \geq 0 \quad (6h)$$

### C. Challenges under the Multi-Region Setting

The above formulation is for a single-region power system. For interconnected multi-region power systems, directly solving the above stochastic ACOPF model is not viable due to the unwillingness of data sharing among different regions. Hence, distributed computation schemes should be leveraged to limit data-sharing activity for better privacy. Another challenge is the quickly increased time cost and the potential divergence issue when the total number of scenarios becomes large. To tackle such challenges, the consensus ADMM algorithm and the K-means-based scenario reduction method are employed in this paper, which will be explained in the next section.

### III. CONSENSUS ADMM AND SCENARIO REDUCTION

### A. Consensus ADMM

The Consensus ADMM is a distributed optimization algorithm well-suited for large-scale and decentralized power systems. The primary goal of the Consensus ADMM is to enable multiple agents (each responsible for a subset of the original problem) to cooperatively solve a reduced-size subproblem via local computations and minimal communications.

#### 1) Fundamentals of ADMM

Traditional ADMM (Alternating Direction Method of Multipliers) is invented by experts to solve decomposable optimization problems. Its basic form is shown in Eq. (7).

$$\min_{x,z} \quad f(x) + g(z) \quad (7a)$$
$$\text{s.t.} \quad Ax + Bz = c \quad (7b)$$

Here, $f(x)$ and $g(z)$ are two parts of the objective function, and $x$ and $z$ are the decision variables. $A$, $B$, and $c$ are parameters in the constraints. Traditional ADMM introduces the Lagrange multipliers, uses alternative optimization to update $x$ and $z$, and then updates the Lagrange multiplier. This process iterates until reaching the stopping criteria.

While traditional ADMM is effective for many engineering problems, it has limitations when applied to distributed ACOPF: it requires significant information exchange among regions, which may incur reluctance from regional entities.

#### 2) Basic Idea of Consensus ADMM

Consensus ADMM ensures that the local solutions from

each region are consistent. The consensus problem can be formulated as:

$$\min_x \sum_{i=1}^{N} f_i(x_i) \quad (8a)$$

$$\text{s.t.} \quad x_i = z, \forall i = 1, ..., N \quad (8b)$$

where $x_i$ represents the local variables of region $i$, and $z$ is the global consensus variable. $N$ is the total number of regions.

The augmented Lagrangian function for the consensus problem is:

$$L_\rho(x_1, ..., x_N, z, \lambda) = \sum_{i=1}^{N} \left( f_i(x_i) + \lambda_i^T(x_i - z) + \frac{\rho}{2} \|x_i - z\|_2^2 \right) \quad (9)$$

where $\lambda_i$ are the Lagrange multipliers, and $\rho$ is the penalty parameter. The Consensus ADMM updates are given by:

$$x_i^{k+1} = \arg\min_{x_i} \left( f_i(x_i) + \lambda_i^T(x_i - z^k) + \frac{\rho}{2} \|x_i - z^k\|_2^2 \right) \text{(in parallel)} \quad (10a)$$

$$z^{k+1} = \frac{1}{N} \sum_{i=1}^{N} x_i^{k+1}, \quad \lambda_i^{k+1} = \lambda_i^k + \rho(x_i^{k+1} - z^{k+1}) \quad (10b)$$

The above $z$-update step can also be computed locally once the updated states from other regions are collected. Hence, the iterations of Consensus ADMM can be conducted in a completely distributed manner, with information exchanged only between the *virtual* consensus agent and each region. Besides, this consensus scheme allows for parallel computing, thus beneficial for interconnected system operation [7][8]. Fig.1 gives an illustration of applying the above idea to the IEEE 14-bus system.

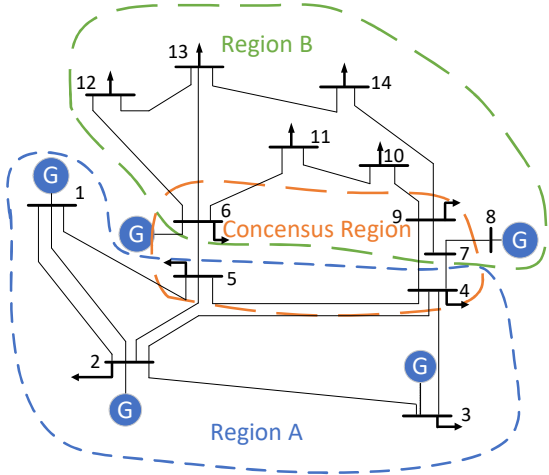

Fig. 1. Two physical regions of the IEEE 14-bus system (top and bottom), plus a virtual consensus region (middle) for ADMM optimization.

### B. Scenario Reduction

As previously discussed, considering massive scenarios can result in the "dimensional exploding" issue for the proposed stochastic ACOPF model. Thus, we employ the following scenario reduction technique to reduce the computational complexity while maintaining accuracy. Based on the reduced scenarios set and the Consensus ADMM, we establish a distributed version of the proposed stochastic ACOPF model.

Here, a combined method of improved K-means clustering and Simultaneous Backward Reduction (SBR) is employed to reduce the load forecasting scenarios. More specifically:

- Improved K-means Clustering [14]: the first cluster center is selected based on the highest density. Subsequent centers maximize the distance from the already chosen centers.

- Simultaneous Backward Reduction (SBR): efficiently reduces the number of scenarios in stochastic programming by iteratively eliminating the least significant scenarios, based on the Kantorovich distance [13], while preserving the overall statistical properties of the original set.

- Scenario Reduction Method: we first apply improved K-means clustering to group load forecasting scenarios, then use the SBR algorithm to reduce the number of scenarios in each cluster. This combined method ensures that the reduced set retains the key characteristics and extremes of the original scenarios, as illustrated in Fig. 2.

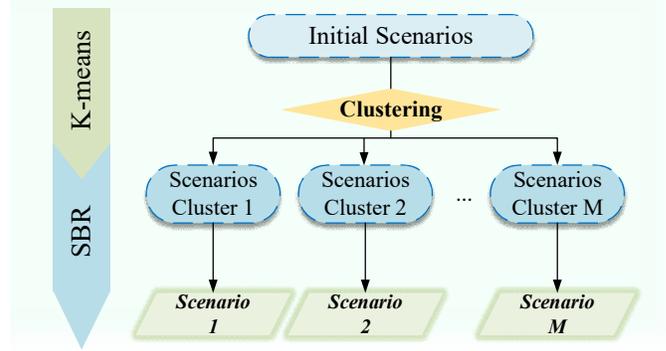

Fig. 2. The process of scenario reduction by using K-means and SBR.

### C. Combined with the Stochastic ACOPF

The Consensus ADMM and scenario reduction are integrated into the stochastic ACOPF model, as described in **Algorithm 1**.

We denote decision variables specific to a particular region using subscripts from the given region segmentation of the power grid. The shared variables between non-consensus regions are handled in a *virtual* consensus agent.

For a region denoted by $\mathcal{A}$, the local variables are defined as:

$$x_\mathcal{A} = \left[ P_{i,\mathcal{A}}, Q_{i,\mathcal{A}}, V_{i,\mathcal{A}}, \theta_{i,\mathcal{A}}, \Delta P_{i,s}^d \right]^T \quad (11)$$

where $P_{i,\mathcal{A}}$ and $Q_{i,\mathcal{A}}$ represent the active and reactive power outputs of generators in the region $\mathcal{A}$ at bus $i$. $V_{i,\mathcal{A}}$ and $\theta_{i,\mathcal{A}}$ are the voltage magnitudes and angles at buses within $\mathcal{A}$.

The consensus region $\mathcal{C}$ (with $m$ buses) holds the variables:

$$z = \left[ V_{1,\mathcal{C}}, ..., V_{m,\mathcal{C}}, \theta_{1,\mathcal{C}}, ..., \theta_{m,\mathcal{C}} \right]^T \quad (12)$$

We define the overlapping parts of $x_\mathcal{A}$ and $z$ as $x_{\mathcal{C},\mathcal{A}}$, which represents the shared variables between the region $\mathcal{A}$ and the consensus region $\mathcal{C}$. Then, the local updates for the region $\mathcal{A}$ at the ($k$+1)-th iteration are given by:

$$x_\mathcal{A}^{k+1} = \arg\min_{x_\mathcal{A}} L_\rho(x_\mathcal{A}, z^k, \lambda_\mathcal{A}^k) \quad (13)$$

where,

$$L_\rho(x_\mathcal{A}, z, \lambda_\mathcal{A}) = f_\mathcal{A}(x_\mathcal{A}) + \lambda_\mathcal{A}^T(x_{\mathcal{CA}} - z) + \frac{\rho}{2}\| x_{\mathcal{CA}} - z \|_2^2 \quad (14)$$

and $f_\mathcal{A}(x_\mathcal{A})$ is the objective function representing generation costs and loss of load costs in the region $\mathcal{A}$. $\lambda_\mathcal{A}^k$ are the Lagrange multipliers associated with the consensus constraints.

The term $\lambda_\mathcal{A}^T(x_{\mathcal{CA}} - z)$ acts as the complementary term, ensuring that the local solution $x_\mathcal{A}^k$ gradually converges to the consensus variable $z^k$. The penalty term $(\rho/2)\|x_{\mathcal{CA}} - z\|_2^2$ penalizes the deviation of $x_{\mathcal{CA}}^k$ from $z^k$, encouraging $x_{\mathcal{CA}}^k$ to approach $z^k$ in each iteration. The larger the value of $\rho$, the stronger the consensus, and the smaller the value of $\rho$, the greater the flexibility of the local solution.

Hence, the optimization problem for the region $\mathcal{A}$ is formulated as follows:

$$\min_{x_\mathcal{A}} L_\rho(x_\mathcal{A}, z, \lambda_\mathcal{A}) \quad (15a)$$

$$\text{s.t.} \quad P_i^{\min} \le P_i \le P_i^{\max}, \forall i \quad (15b)$$

$$Q_i^{\min} \le Q_i \le Q_i^{\max}, \forall i \quad (15c)$$

$$V_j^{\min} \le |V_j| \le V_j^{\max}, \forall j \quad (15d)$$

$$\theta_j^{\min} \le \theta_j \le \theta_j^{\max}, \forall j \quad (15d)$$

$$0 \le \Delta P_{i,s}^d \le P_{i,s}^d, \forall i, \forall s \quad (15f)$$

$$Q_i - Q_i^d - \sum_j V_i V_j (G_{ij} \sin(\theta_i - \theta_j) - B_{ij} \cos(\theta_i - \theta_j)) = 0, \forall i, \forall j \quad (15g)$$

$$\Delta P_{i,s}^d + P_i - P_{i,s}^d - \sum_j V_i V_j (G_{ij} \cos(\theta_i - \theta_j) + B_{ij} \sin(\theta_i - \theta_j)) \ge 0, \forall i, \forall j, \forall s \quad (15h)$$

where $i$ is a bus-index within the region $\mathcal{A}$, $j$ is a bus-index within the union of regions $\mathcal{A}$ and $\mathcal{C}$. $s$ stands for a load forecasting scenario in the reduced scenario set $\mathcal{S}^*$.

By solving their respective stochastic ACOPF problems (i.e., Eq. (15)), all non-consensus regions can work in parallel, as shown in **Algorithm 1**.

---

**Algorithm 1** Consensus ADMM Algorithm with Scenario Reduction

Input original scenario set $\mathcal{S}$, perform scenario reduction algorithm in Fig.2 to obtain reduced set of scenarios $\mathcal{S}^*$ and probability $\mu^*$

Initialize $\rho$, tolerance and decision variables: $P_\mathcal{A}, Q_\mathcal{A}, V_\mathcal{A}, \theta_\mathcal{A}, \Delta P_{\mathcal{A}i,s}^d$
$P_\mathcal{B}, Q_\mathcal{B}, V_\mathcal{B}, \theta_\mathcal{B}, \Delta P_{\mathcal{B}i,s}^d$

Initialize consensus variable $z$ and Lagrange multipliers $\lambda_\mathcal{A}$ and $\lambda_\mathcal{B}$

Define objective functions $obj_\mathcal{A}$ and $obj_\mathcal{B}$ and constraints $st_\mathcal{A}$ and $st_\mathcal{B}$ for region $\mathcal{A}$ and region $\mathcal{B}$

**while** $error \ge tolerance$ **do**

    **Parallel processing:**

        Solve the optimization problem for region $\mathcal{A}$ to obtain $x_\mathcal{A}^k$

        Solve the optimization problem for region $\mathcal{B}$ to obtain $x_\mathcal{B}^k$

    Compute consensus variables $x_{\mathcal{CA}}^k$ and $x_{\mathcal{CB}}^k$ from $x_\mathcal{A}^k$ and $x_\mathcal{B}^k$

    Update $z^{k+1} \leftarrow \frac{1}{2}(x_{\mathcal{CA}}^k + x_{\mathcal{CB}}^k)$

    Update $\lambda_\mathcal{A}^{k+1} \leftarrow \lambda_\mathcal{A}^k + \rho(x_{\mathcal{CA}}^k - z^{k+1})$ and $\lambda_\mathcal{B}^{k+1} \leftarrow \lambda_\mathcal{B}^k + \rho(x_{\mathcal{CB}}^k - z^{k+1})$

    Update consensus variable $z \leftarrow z^{k+1}$

    Compute $error \leftarrow \|z^{k+1} - z^k\|$

**end while**

Output the optimal results

---

## IV. CASE STUDY

This section presents two case studies investigating the proposed stochastic ACOPF model and the Consensus ADMM approach. The first case study uses the IEEE 14-bus system, and the second uses the IEEE 30-bus test system. The result's optimality and feasibility are compared with the baseline model.

### A. Experiment Setup and Performance Metrics

In the experiments here, the penalty cost for load loss was set to ten times the unit generation cost, providing a significant incentive to avoid load shedding. The penalty parameter $\rho$ was uniformly set at $10^6$ across all experiments. The original load and system data for these case studies are from [16]. Initially, 100 load scenarios are sampled from a Gaussian distribution [9] $N(P_d, (0.1P_d)^2)$, where $P^d$ stands for the original load. All the proposed optimization models are implemented in MATLAB 2023b and solved by lower-level interior-point algorithms [17]. The experiments are conducted on a machine with an AMD Ryzen 7 8845HS CPU and 16GB RAM.

To evaluate the solution quality under *all original* scenarios, the following reliability index is used, called "loss of slack-power probability (**LOSP**)" or "slack-power shortage rate":

$$\mathbf{E}_\mathbf{S}\left[\mathbf{1}\left(P_{slack}^{OPF} < P_{slack}^{basePF}\right)\right] \approx \frac{1}{M}\sum_{s=1}^{M}\mathbf{1}\left(P_{slack}^{OPF} < P_{slack,s}^{basePF}\right) \quad (16)$$

where, $P_{slack}^{OPF}$ is the active power of the slack bus from solving a specific OPF model. $P_{slack,s}^{basePF}$ is the active power of the slack bus from solving the *usual* power flow problem under the $s$-th load scenario after substituting the baseline OPF solution for generator buses' setting points (i.e., $P_G$ and $V_G$ of the PV-type buses). $M$ is the total number of original (i.e., unreduced) scenarios. $\mathbf{E}_\mathbf{S}$ is the expectation operator w.r.t the load scenario-related random variable $\mathbf{S}$. $\mathbf{1}$ is the "indicator" function (equals one if the event holds; otherwise zero). The LOSP characterizes the expected extent of power shortage since one interpretation of the "slack-bus power" (when positive) is the power imported from the external grid. A larger LOSP means a larger possibility of asking for external support. Therefore, it can be adopted as a systemwide reliability index.

### B. Case Study on IEEE 14-bus System

The IEEE 14-bus system consists of 14 buses, 20 branches, and five generators. The comparison results are listed in Table I.

TABLE I. COMPARISON RESULTS OF DIFFERENT MODELS APPLIED TO THE IEEE 14-BUS SYSTEM

| Methods | Baseline | ADMM | Baseline (Stochastic) | ADMM (Stochastic) |
|---|---|---|---|---|
| Total cost ($/hr) | 8081.53 | 8036.00 | 9532.62 | **9238.09** |
| LOSP | - | - | 51% | **3%** |
| $P_1$ (MW) | 194.33 | 177.93 | 194.33 | 179.55 |
| $Q_1$ (MVar) | 0.00 | 2.24 | 0.00 | 0.03 |
| $P_2$ (MW) | 36.72 | 33.94 | 36.72 | 34.26 |
| $Q_2$ (MVar) | 23.69 | -2.30 | 23.69 | 1.40 |
| $P_3$ (MW) | 28.74 | 0.00 | 28.74 | 8.97 |
| $Q_3$ (MVar) | 24.13 | 19.69 | 24.13 | 18.73 |
| $P_6$ (MW) | 0.00 | 52.98 | 0.00 | 65.31 |
| $Q_6$ (MVar) | 11.55 | 24.00 | 11.55 | 23.99 |
| $P_8$ (MW) | 8.49 | 0.00 | 8.49 | 0.00 |
| $Q_8$ (MVar) | 8.27 | 24.00 | 8.27 | 23.99 |

In Table I, the "Baseline" column presents results from solving the standard ACOPF model by the conventional algorithm on the (original) base case data in MATPOWER (i.e., not considering the load stochasticity). Similarly, the "ADMM" column gives results from solving the standard ACOPF model by the consensus ADMM algorithm on the base case data.

The "Baseline (Stochastic)" column presents results from substituting the baseline OPF solution into the objective function to calculate the averaged (i.e., expected) total cost considering load loss. Then, the baseline OPF solution is substituted into Eq. (16) to obtain the slack-power shortage rate. This approach yields a significantly higher cost of $9532.62/hr.

The "ADMM (Stochastic)" column presents results from solving the proposed distributed stochastic ACOPF model (with load loss cost) under the *reduced* scenarios. Then, substituting its solution into Eq. (16) to calculate the slack-power shortage rate. It achieves a lower cost of $9238.09/hr and a lower slack-power shortage rate than the "Baseline (Stochastic)" column.

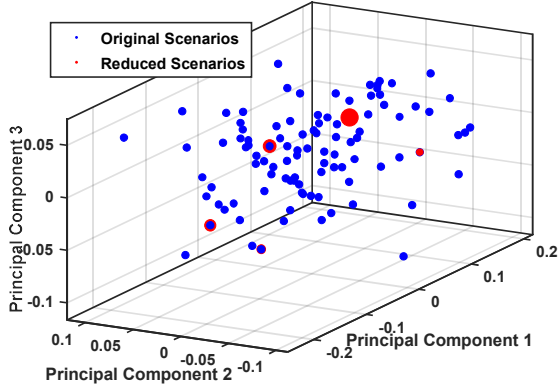

Fig. 3. The original and the reduced load forecasting scenarios (case study I).

The original and the reduced load forecasting scenarios are visualized in Fig. 3 using Principal Component Analysis (PCA) for 2D-display. The reduced scenarios, obtained by combining K-means clustering and SBR methods, are marked in red. The size of the reduced scenario points indicates their probability.

Fig. 4 depicts the voltage magnitudes $V$ and voltage angles $\theta$ for regions $\mathcal{A}$ and $\mathcal{B}$ over iterations. The subplots $V_\mathcal{A}$, $V_\mathcal{B}$, $\theta_\mathcal{A}$ and $\theta_\mathcal{B}$ show how these variables converge to stable values as the iteration proceeds, demonstrating the capability of the consensus ADMM to achieve an equilibrium between the two regions in a distributed manner.

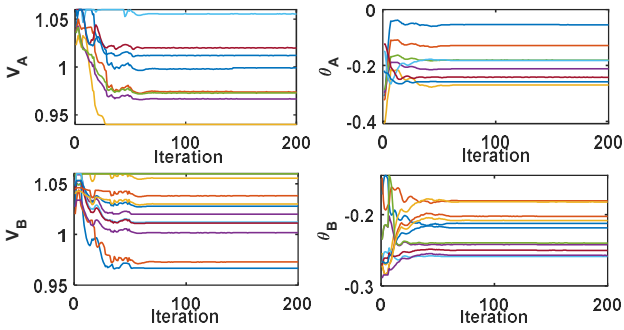

Fig. 4. $V_\mathcal{A}$, $V_\mathcal{B}$ and $\theta_\mathcal{A}$, $\theta_\mathcal{B}$ for regions $\mathcal{A}$ and $\mathcal{B}$ over iterations.

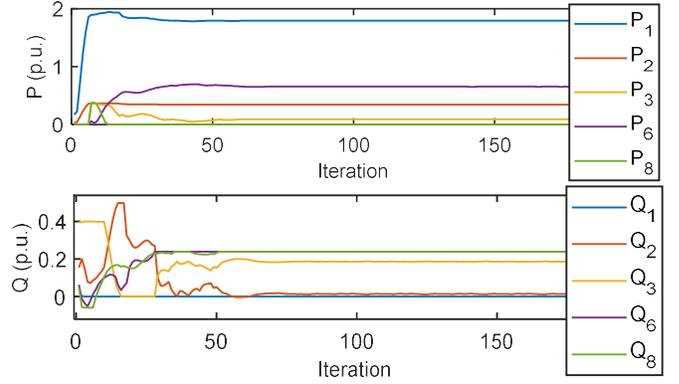

Fig. 5. Generator outputs over iterations.

Fig. 5 depicts the active power outputs $P_i$ and reactive power outputs $Q_i$ of the generators over the iterations.

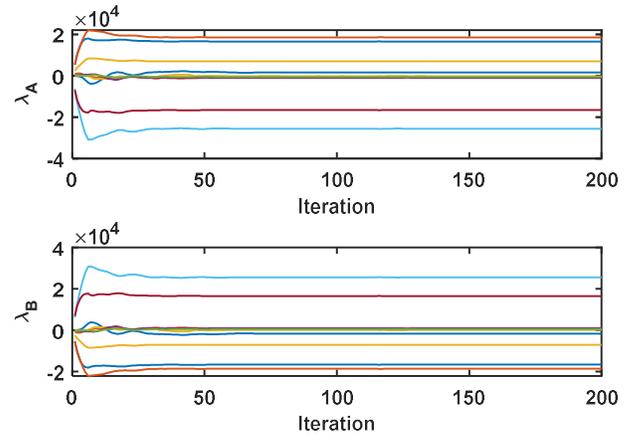

Fig. 6. Convergence of the Lagrange multipliers for regions $\mathcal{A}$ and $\mathcal{B}$.

Fig. 6 displays how the Lagrange multipliers (representing the shadow cost associated with the consensus constraints) gradually stabilize as the iteration proceeds.

### C. Case Study on IEEE 30-bus System

The IEEE 30-bus system represents a simplified network of the AEP system in the early 1960s. It consists of 30 buses, 41 branches, and six generators.

TABLE II. COMPARISON RESULTS FOR DIFFERENT METHODS APPLIED TO THE IEEE 30-BUS SYSTEM

| Methods | Baseline | ADMM | Baseline (Stochastic) | ADMM (Stochastic) |
|---|---|---|---|---|
| Total cost ($/hr) | 576.89 | 557.28 | 646.89 | **635.64** |
| LOSP | - | - | 52% | **2%** |
| $P_1$ (MW) | 41.54 | 21.46 | 41.54 | 57.00 |
| $Q_1$ (MVar) | 35.93 | 33.22 | 35.93 | 7.39 |
| $P_2$ (MW) | 55.40 | 31.34 | 55.40 | 22.36 |
| $Q_2$ (MVar) | 34.20 | 48.70 | 34.20 | 31.25 |
| $P_{13}$ (MW) | 16.20 | 16.79 | 16.20 | 39.64 |
| $Q_{13}$ (MVar) | 6.96 | -10.00 | 6.96 | 48.70 |
| $P_{22}$ (MW) | 22.74 | 17.04 | 22.74 | 19.23 |
| $Q_{22}$ (MVar) | 31.75 | -10.62 | 31.75 | -10.00 |
| $P_{23}$ (MW) | 16.27 | 46.46 | 16.27 | 19.69 |
| $Q_{23}$ (MVar) | -5.44 | 25.51 | -5.44 | -10.88 |
| $P_{27}$ (MW) | 39.91 | 53.88 | 39.91 | 49.43 |
| $Q_{27}$ (MVar) | 1.67 | 6.92 | 1.67 | 27.42 |

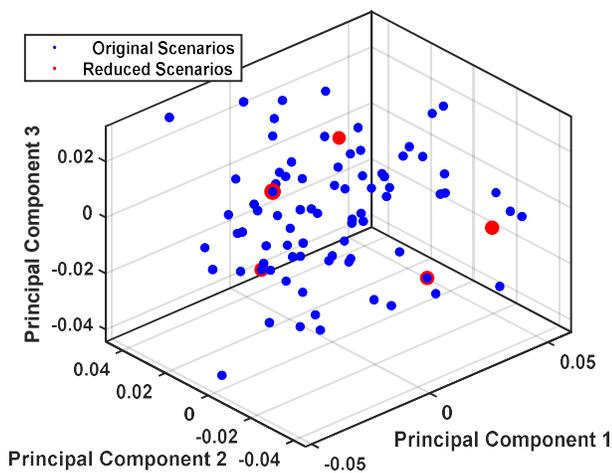

Fig. 7. The original and the reduced load forecasting scenarios (case study II).

Like the previous case study, the original and the reduced load forecasting scenarios are depicted in Fig. 7. TABLE II lists the total cost, slack-power shortage rate, and generation outputs from all four approaches. The results of the first two rows again demonstrate the Consensus ADMM's capability to tackle the proposed stochastic ACOPF problem.

## V. Conclusion

In this paper, we presented a new approach for modeling and solving the stochastic ACOPF based on Consensus ADMM. By integrating scenario reduction techniques, our approach is scalable for power systems of different sizes. The proposed approach is tested on IEEE 14-bus and IEEE 30-bus systems, demonstrating an observable improvement in system reliability index and cost-reduction under stochastic load scenarios. This approach can be applied in the generation planning of large interconnected power systems.

Future work includes 1) extending this approach to more complicated power systems, e.g., the 118-bus system, and 2) considering renewable sources (such as solar and wind) and their stochasticity in the proposed distributed ACOPF model.

## References

[1] M. Alkhraijah, R. Harris, C. Coffrin and D. K. Molzahn, "PowerModelsADA: A Framework for Solving Optimal Power Flow Using Distributed Algorithms," in *IEEE Transactions on Power Systems*, vol. 39, no. 1, pp. 2357-2360, Jan. 2024.

[2] K. Sun and X. A. Sun, "A Two-Level ADMM Algorithm for AC OPF With Global Convergence Guarantees," in IEEE Transactions on Power Systems, vol. 36, no. 6, pp. 5271-5281, Nov. 2021.

[3] C. Gavriluta, R. Caire, A. Gomez-Exposito and N. Hadjsaid, "A Distributed Approach for OPF-Based Secondary Control of MTDC Systems," in *IEEE Transactions on Smart Grid*, vol. 9, no. 4, pp. 2843-2851, July 2018.

[4] J. Guo, G. Hug and O. K. Tonguz, "A Case for Nonconvex Distributed Optimization in Large-Scale Power Systems," in *IEEE Transactions on Power Systems*, vol. 32, no. 5, pp. 3842-3851, Sept. 2017.

[5] T. Erseghe, "Distributed Optimal Power Flow Using ADMM," in *IEEE Transactions on Power Systems*, vol. 29, no. 5, pp. 2370-2380, Sept. 2014.

[6] S. Boyd, N. Parikh, E. Chu, B. Peleato, and J. Eckstein. "Distributed Optimization and Statistical Learning via the Alternating Direction Method of Multipliers," in *Foundations and Trends in Machine Learning*, 3(1):1-122, 2011.

[7] Q. Li, B. Kailkhura, R. Goldhahn, P. Ray and P. K. Varshney, "Robust Decentralized Learning Using ADMM With Unreliable Agents," in *IEEE Transactions on Signal Processing*, vol. 70, pp. 2743-2757, 2022.

[8] R. S. Kar, Z. Miao, M. Zhang and L. Fan, "ADMM for nonconvex AC optimal power flow," 2017 North American Power Symposium (NAPS), Morgantown, WV, USA, 2017, pp. 1-6.

[9] L. Xiao-fei and S. Li-qun, "Power system load forecasting by improved principal component analysis and neural network," 2016 IEEE International Conference on High Voltage Engineering and Application (ICHVE), Chengdu, China, 2016, pp. 1-4.

[10] R Y. Li, W. Li, W. Yan, J. Yu and X. Zhao, "Probabilistic Optimal Power Flow Considering Correlations of Wind Speeds Following Different Distributions," in *IEEE Transactions on Power Systems*, vol. 29, no. 4, pp. 1847-1854, July 2014.

[11] X. Li, Y. Li and S. Zhang, "Analysis of Probabilistic Optimal Power Flow Taking Account of the Variation of Load Power," in *IEEE Transactions on Power Systems*, vol. 23, no. 3, pp. 992-999, Aug. 2008.

[12] Taiyou Yong and R. H. Lasseter, "Stochastic optimal power flow: formulation and solution," 2000 Power Engineering Society Summer Meeting (Cat. No.00CH37134), Seattle, WA, USA, 2000, pp. 237-242 vol. 1, doi: 10.1109/PESS.2000.867521.

[13] M. Huang, X. Zheng, Z. Liao and X. Huang, "Modeling and Analysis for Power Substation Load Data based on Spectral Clustering," 2021 IEEE 4th International Electrical and Energy Conference (CIEEC), Wuhan, China, 2021, pp. 1-4, doi: 10.1109/CIEEC50170.2021.9510616.

[14] J. M. Morales, S. Pineda, A. J. Conejo and M. Carrion, "Scenario Reduction for Futures Market Trading in Electricity Markets," in *IEEE Transactions on Power Systems*, vol. 24, no. 2, pp. 878-888, May 2009.

[15] Y. Shi, S. Gao and X. Zhao, "Extraction of typical scenarios of power grid operation based on improved K-means clustering algorithm," 2021 IEEE Sustainable Power and Energy Conference (iSPEC), Nanjing, China, 2021, pp. 2628-2633, doi: 10.1109/iSPEC53008.2021.9736128.

[16] R. D. Zimmerman, C. E. Murillo-Sánchez and R. J. Thomas, "MATPOWER: Steady-State Operations, Planning, and Analysis Tools for Power Systems Research and Education," in *IEEE Transactions on Power Systems*, vol. 26, no. 1, pp. 12-19, Feb. 2011.

[17] Nocedal, Jorge, Figen Öztoprak, and Richard A. Waltz. An Interior Point Method for Nonlinear Programming with Infeasibility Detection Capabilities. *Optimization Methods & Software* 29(4), July 2014, pp. 837–854.